\documentclass[prl,superscriptaddress,twocolumn,preprintnumbers,a4,showpacs,floatfix]{revtex4}  
\usepackage{amsmath, amssymb, psfrag, tabls, dcolumn, bm, color}
\usepackage[sort&compress]{natbib}
\usepackage[pdftex]{graphicx}
\usepackage{graphicx}

\begin{document}

\title{Evidence for Graphene Edges Beyond Zigzag and Armchair}

\author{Pekka Koskinen\footnote{Author to whom correspondence should be addressed.}}
\affiliation{Department of Physics, NanoScience Center, 40014 University of Jyv\"askyl\"a, Finland}
\email{pekka.koskinen@iki.fi}

\author{Sami Malola}
\affiliation{Department of Physics, NanoScience Center, 40014 University of Jyv\"askyl\"a, Finland}

\author{Hannu H\"akkinen}
\affiliation{Department of Physics, NanoScience Center, 40014 University of Jyv\"askyl\"a, Finland}
\affiliation{Department of Chemistry, NanoScience Center, 40014 University of Jyv\"askyl\"a, Finland}


%
%
%
%

\begin{abstract}
The edges of nanoscopic objects determine most of their properties. For this reason the edges of honeycomb carbon---always considered either zigzag- or armchair-like---need special attention. In this report we provide experimental evidence confirming a previous unexpected prediction: zigzag is a metastable edge, as its planar reconstruction lowers energy and forms the most stable graphene edge. Our evidence is based on re-analyzing a recent experiment. Since the reconstructed edge, along with other unconventional edges we discuss, has distinct chemical properties, this discovery urges for care in experiments and theory---we must enter the realm beyond zigzag and armchair.
\end{abstract}

\maketitle
\hfill

Graphene is a two-dimensional sheet of carbon atoms with the honeycomb structure\cite{novoselov_science_04}, and it underlies other carbon allotropes like graphite, carbon nanotubes, and fullerenes\cite{geim_nmat_07}. If you cut honeycomb carbon in a random orientation, the edges become purely zigzag- or armchair-like, or an alternating series of zigzag and armchair segments. So far these edges have been taken for granted, both in experiment and in theory.

The edges of graphene are prominent for several reasons. For example, carbon nanotubes are made from curved graphene sheets, and the edges determine how they grow\cite{lee_PRL_97,loginova_NJP_08} or make contacts\cite{jian_carbon_06}. Furthermore, conductive\cite{jia_science_09}, mechanical and elastic properties\cite{jun_PRB_08,malola_EPJD_09,bhuang_PRL_09}, as well as chemical properties\cite{jiang_JCP_07} of graphene nanoribbons depend crucially on the edge properties (edge profile reflects the symmetry inside ribbons). One example is the electronic edge state at the zigzag edge, completely missing from the armchair edge\cite{kobayashi_PRB_06}. 

\begin{figure}[t!]
\includegraphics[width=9cm]{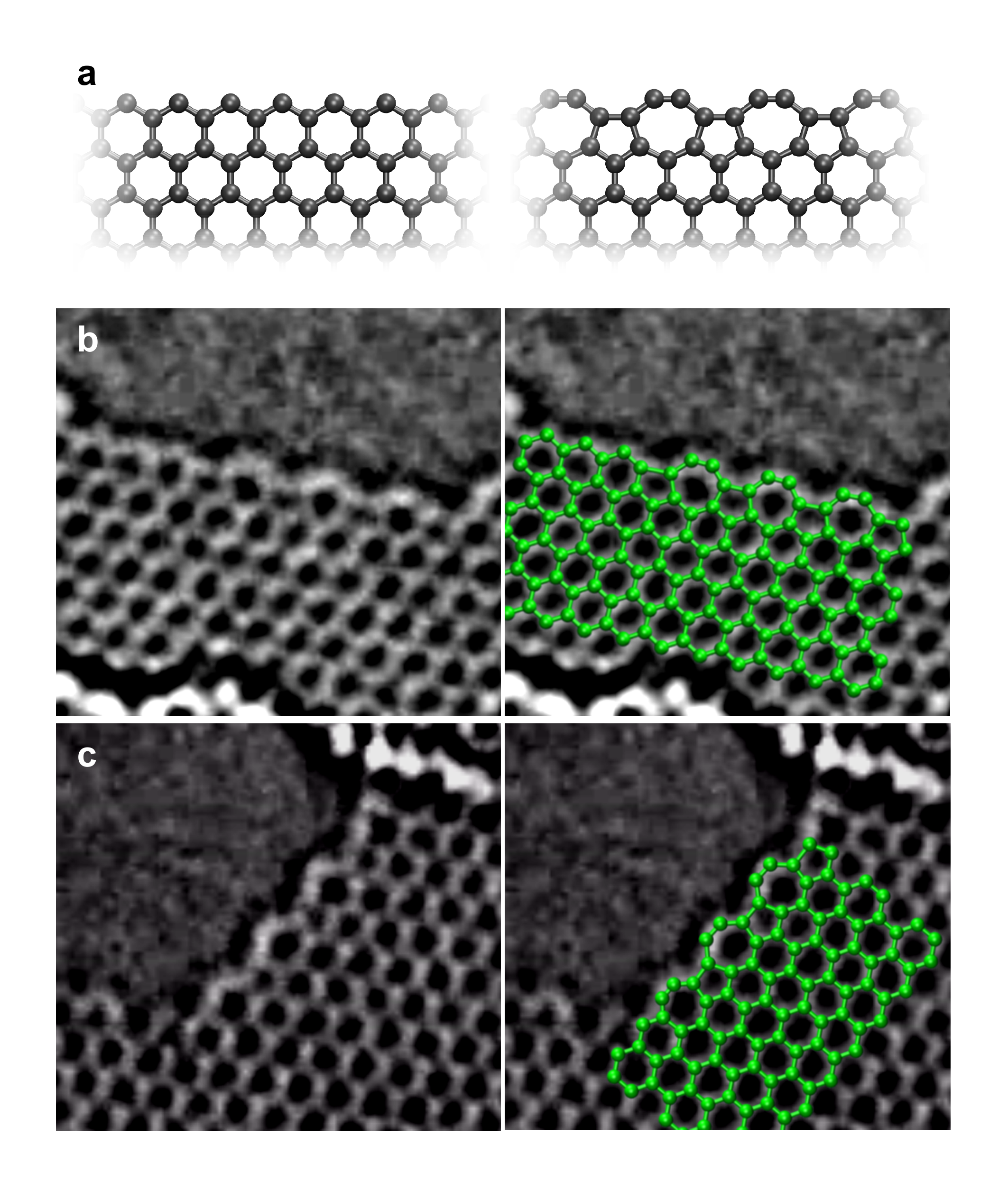}
\caption{(color online) Experimental evidence for zigzag reconstruction. (a) Normal zigzag edge (left) and the reconstructed zigzag edge (right), as predicted in Ref.~\onlinecite{koskinen_PRL_08} (b) Frame number $103$ from the Movie S1 of Ref.~\onlinecite{girit_science_09}, showing the zigzag edge reconstruction (left), with a highlighted structure assignment (right). Note the transition from pure zigzag to reconstructed zigzag edge. (c) Frame number $64$ from the same movie with another reconstructed zigzag edge (left), and a structure assignment (right). Printed with the kind permission of authors of Ref.~\onlinecite{girit_science_09}.}
\label{fig1}
\end{figure}

In a recent theoretical work Koskinen \emph{et al.} considered also other than pure zigzag or armchair edges\cite{koskinen_PRL_08}. Computer simulations using density-functional theory predicted, among other results, the existence of a reconstructed zigzag edge, shown in Fig.~\ref{fig1}a. The energy of the edge is lowered by $0.35$~eV/\AA\ ($1.7$~eV per two adjacent hexagons) when zigzag reconstructs; this makes the reconstructed zigzag the most stable graphene edge. These results have later been confirmed in subsequent theoretical work\cite{wassmann_PRL_08,bhuang_PRL_09,reddy_APL_09}. Furthermore, besides lowering the energy, the reconstruction changes the edges' chemical properties. The strong dangling bonds, responsible for the reactivity of the zigzag edge, are removed by the reconstruction due to the formation of triple-bonds in the nearly linear armrest parts\cite{kawai_PRB_00} (cf. Fig.~\ref{fig1}a). Up to now this edge has been an academic curiosity, but here we present experimental evidence confirming the prediction: zigzag and armchair edges do have company.

A recent impressive experiment by Girit~\emph{et al.} (``the experiment'' from now on) imaged the edges of a free-standing graphene by aberration-corrected transmission electron microscopy (TEM)\cite{girit_science_09}. Sub-\AA ngstr\"om resolution enabled the analysis of edge structures and dynamics with atom precision. The movement of atoms at the edge, driven by collisions with accelerated electrons, could be understood by simple arguments. The experimental report was supplemented by a movie (Movie S1) that animated the edge dynamics with $110$ frames corresponding to some $10$ minutes of real time ($1$~second exposure for each imaged frame with $4$~second breaks). Figs.~\ref{fig1}b and \ref{fig1}c show two frames where the reconstructed zigzag edges are apparent (such frames were easy to find). The analysis in the original experimental report considered only zigzag and armchair edges because they contain regular hexagons; other polygons were rejected since they gave merely a messy appearance for the edges. This is natural, because---until now---no other edge structures have been recognized in an experiment; any deviations have been viewed as \emph{local} defects. But since now Fig.~\ref{fig1}a gives a fresh viewpoint for the experiment, it becomes evident that the reconstructed zigzag edge is a well-defined, stable, and proper edge structure in itself. On the account of this experimental discovery, we term the reconstructed zigzag edge the \emph{reczag} edge. In the experiment reczag edges appear with fair abundance, and they sometimes persist for well more than ten seconds in real-time (several frames).

But how can we tell that zigzag really is metastable? First of all, the energy difference between zigzag and reczag edges is large enough to be theoretically beyond any doubts---from that standpoint the reczag edge has been nothing but waiting for discovery\cite{koskinen_PRL_08}. But also experimental considerations support metastability. As carefully explained in the experimental article, zigzag edges appear more abundant than armchair edges due to dynamic and kinematic effects related to TEM imaging, not due to lower energy\cite{girit_science_09}. In short, when TEM-accelerated electrons cause damage and kick doubly coordinated atoms at the edges, the zigzag edges are repaired rapidly by diffusing single carbon atoms, while for armchair the repair requires two atoms, hence it is slower. The reczag edge has an edge profile similar to armchair, and is therefore also discriminated by the TEM imaging process. Furthermore, in the experiment the edge is eroded so that graphene with pristine hexagons is revealed during the course of time, favoring zigzag instead of reczag edge (that contains other polygons). Regardless of these discriminating factors, fairly long segments of the reczag edges can be observed.

\begin{figure}
\includegraphics[width=8cm]{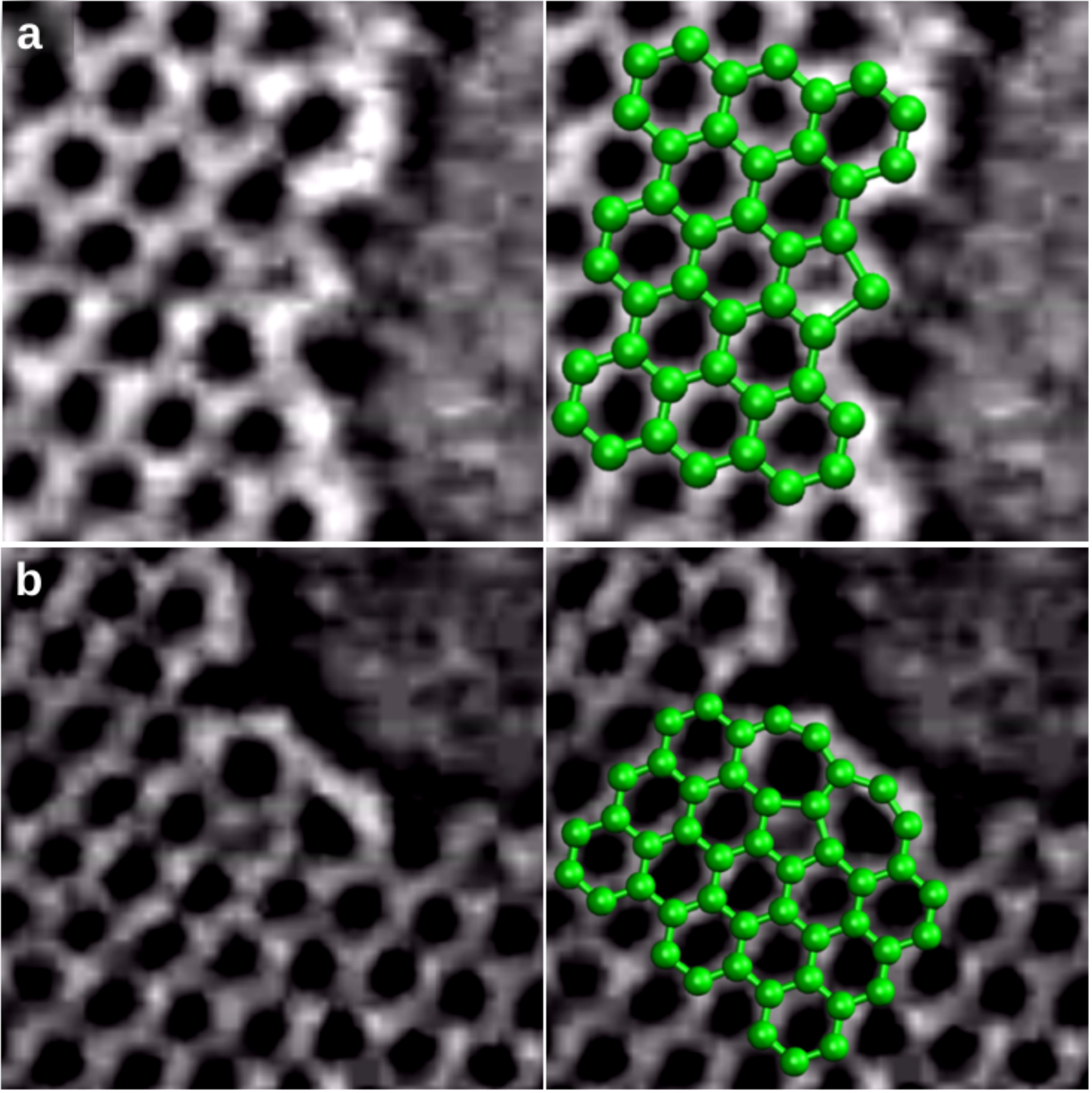}
\caption{(color online) Segments of other graphene edges, as seen from frame details of the Movie S1 of Ref.~\onlinecite{girit_science_09}. (a) Detail of frame $73$ (left) with structure assignment (right), revealing an armchair edge with pentagons. (b) Detail of frame $106$ (left) with structure assignment (right), revealing an armchair edge with hexagons and two heptagons. Printed with the kind permission of authors of Ref.~\onlinecite{girit_science_09}. }
\label{fig2}
\end{figure}

A systematic inspection of the experimental data reveals short segments of the other predicted edges as well, such as an armchair edge with pentagons in Fig.~\ref{fig2}a, or an armchair edge with two adjacent heptagons in Fig.~\ref{fig2}b~\cite{koskinen_PRL_08}. In fact, Ref.~\onlinecite{bhuang_PRL_09} predicted a partial metastability also for the armchair edge: it can relieve edge stress by forming heptagons at the edge and hereby decrease energy, attaining energy minimum with $\sim 30$~\% heptagon concentration. However, since there are no long segments of armchair edges in the experiment, the edge stress does not need any relief, and heptagons at the armchair edge are only rarely observed (the energy difference is small, too).

An interesting experimental detail is the brightness in the armrest parts in armchair and reczag edges. It is due to the larger electron density between the armrest atoms, in chemical triple bonds---observed directly by TEM\cite{kawai_PRB_00}.

This Brief Report has a brief message: \emph{the reczag edge really exists---and makes a difference}. In current frantic graphene research this is important to digest. First, less edge stress but more mechanical stiffness induces different warping and curling properties for reczag.\cite{bhuang_PRL_09,shenoy_PRL_08,malola_EPJD_09} Second, reczag has vibrational properties akin to those of armchair; this is a dangerous origin for wrong identification\cite{malola_EPJD_09,lan_PRB_09}. Third, while zigzag's the so-called edge-state survives the reconstruction (because its origin is bulk, not edge locally), the dangling bond band in reczag moves away from the Fermi-level, trembling edge's conductive properties.\cite{koskinen_PRL_08} Fourth, and most important, reczag is chemically less reactive than zigzag.\cite{koskinen_PRL_08} A hydrogen-passivated zigzag edge, on the other hand, is more stable than hydrogen-passivated reczag edge.\cite{koskinen_PRL_08,wassmann_PRL_08}

Note also that, unfortunately, graphene edges cannot be identified from the bulk lattice orientation anymore.

The observations in this report are trivial, but they have solid significance. Several properties of reczag, as listed above, are different from the properties of zigzag, yielding different interactions and signatures at the edge. Hence the reczag edge, now with an experimental verification, has to be considered---even re-considered---in any theoretical or experimental work involving honeycomb carbon.

\emph{Note added in proof:} Another experiment (Chuvilin \emph{et al.}, arXiv:0905.3090) has also confirmed our observations.

We acknowledge the Academy of Finland for funding and the authors of Ref.~\onlinecite{girit_science_09} for their kind permission to print their data. Figures reprinted with permission from AAAS.


\begin{thebibliography}{18}
\expandafter\ifx\csname natexlab\endcsname\relax\def\natexlab#1{#1}\fi
\expandafter\ifx\csname bibnamefont\endcsname\relax
  \def\bibnamefont#1{#1}\fi
\expandafter\ifx\csname bibfnamefont\endcsname\relax
  \def\bibfnamefont#1{#1}\fi
\expandafter\ifx\csname citenamefont\endcsname\relax
  \def\citenamefont#1{#1}\fi
\expandafter\ifx\csname url\endcsname\relax
  \def\url#1{\texttt{#1}}\fi
\expandafter\ifx\csname urlprefix\endcsname\relax\def\urlprefix{URL }\fi
\providecommand{\bibinfo}[2]{#2}
\providecommand{\eprint}[2][]{\url{#2}}

\bibitem[{\citenamefont{Novoselov et~al.}(2004)\citenamefont{Novoselov, Geim,
  Morozov, Jiang, Zhang, Dubonos, Grigorieva, and
  Firsov}}]{novoselov_science_04}
\bibinfo{author}{\bibfnamefont{K.~S.} \bibnamefont{Novoselov}},
  \bibinfo{author}{\bibfnamefont{A.~K.} \bibnamefont{Geim}},
  \bibinfo{author}{\bibfnamefont{S.~V.} \bibnamefont{Morozov}},
  \bibinfo{author}{\bibfnamefont{D.}~\bibnamefont{Jiang}},
  \bibinfo{author}{\bibfnamefont{Y.}~\bibnamefont{Zhang}},
  \bibinfo{author}{\bibfnamefont{S.~V.} \bibnamefont{Dubonos}},
  \bibinfo{author}{\bibfnamefont{I.~V.} \bibnamefont{Grigorieva}},
  \bibnamefont{and} \bibinfo{author}{\bibfnamefont{A.~A.}
  \bibnamefont{Firsov}}, \bibinfo{journal}{Science}
  \textbf{\bibinfo{volume}{306}}, \bibinfo{pages}{666} (\bibinfo{year}{2004}).

\bibitem[{\citenamefont{Geim and Novoselov}(2007)}]{geim_nmat_07}
\bibinfo{author}{\bibfnamefont{A.~K.} \bibnamefont{Geim}} \bibnamefont{and}
  \bibinfo{author}{\bibfnamefont{K.~S.} \bibnamefont{Novoselov}},
  \bibinfo{journal}{Nature materials} \textbf{\bibinfo{volume}{6}},
  \bibinfo{pages}{183} (\bibinfo{year}{2007}).

\bibitem[{\citenamefont{Lee et~al.}(1997)\citenamefont{Lee, Kim, and
  Tomanek}}]{lee_PRL_97}
\bibinfo{author}{\bibfnamefont{Y.~H.} \bibnamefont{Lee}},
  \bibinfo{author}{\bibfnamefont{S.~G.} \bibnamefont{Kim}}, \bibnamefont{and}
  \bibinfo{author}{\bibfnamefont{D.}~\bibnamefont{Tomanek}},
  \bibinfo{journal}{Phys. Rev. Lett.} \textbf{\bibinfo{volume}{78}},
  \bibinfo{pages}{2393} (\bibinfo{year}{1997}).

\bibitem[{\citenamefont{Loginova et~al.}(2008)\citenamefont{Loginova, Bartelt,
  Feibelman, and McCarty}}]{loginova_NJP_08}
\bibinfo{author}{\bibfnamefont{E.}~\bibnamefont{Loginova}},
  \bibinfo{author}{\bibfnamefont{N.~C.} \bibnamefont{Bartelt}},
  \bibinfo{author}{\bibfnamefont{P.~J.} \bibnamefont{Feibelman}},
  \bibnamefont{and} \bibinfo{author}{\bibfnamefont{K.~F.}
  \bibnamefont{McCarty}}, \bibinfo{journal}{New J. Phys.}
  \textbf{\bibinfo{volume}{10}}, \bibinfo{pages}{093026}
  (\bibinfo{year}{2008}).

\bibitem[{\citenamefont{Jian et~al.}(2006)\citenamefont{Jian, Yan, Kulaots,
  Crawford, and Hurt}}]{jian_carbon_06}
\bibinfo{author}{\bibfnamefont{K.}~\bibnamefont{Jian}},
  \bibinfo{author}{\bibfnamefont{A.}~\bibnamefont{Yan}},
  \bibinfo{author}{\bibfnamefont{I.}~\bibnamefont{Kulaots}},
  \bibinfo{author}{\bibfnamefont{G.~P.} \bibnamefont{Crawford}},
  \bibnamefont{and} \bibinfo{author}{\bibfnamefont{R.}~\bibnamefont{Hurt}},
  \bibinfo{journal}{Carbon} \textbf{\bibinfo{volume}{44}},
  \bibinfo{pages}{2102} (\bibinfo{year}{2006}).

\bibitem[{\citenamefont{Jia et~al.}(2009)\citenamefont{Jia, Hofmann, Meunier,
  Sumpter, Campos-Delgado, Romo-Herrera, Son, Hsieh, Reina, Kong
  et~al.}}]{jia_science_09}
\bibinfo{author}{\bibfnamefont{X.}~\bibnamefont{Jia}},
  \bibinfo{author}{\bibfnamefont{M.}~\bibnamefont{Hofmann}},
  \bibinfo{author}{\bibfnamefont{V.}~\bibnamefont{Meunier}},
  \bibinfo{author}{\bibfnamefont{B.~G.} \bibnamefont{Sumpter}},
  \bibinfo{author}{\bibfnamefont{J.}~\bibnamefont{Campos-Delgado}},
  \bibinfo{author}{\bibfnamefont{J.~M.} \bibnamefont{Romo-Herrera}},
  \bibinfo{author}{\bibfnamefont{H.}~\bibnamefont{Son}},
  \bibinfo{author}{\bibfnamefont{Y.-P.} \bibnamefont{Hsieh}},
  \bibinfo{author}{\bibfnamefont{A.}~\bibnamefont{Reina}},
  \bibinfo{author}{\bibfnamefont{J.}~\bibnamefont{Kong}}, \bibnamefont{et~al.},
  \bibinfo{journal}{Science} \textbf{\bibinfo{volume}{323}},
  \bibinfo{pages}{1701} (\bibinfo{year}{2009}).

\bibitem[{\citenamefont{Jun}(2008)}]{jun_PRB_08}
\bibinfo{author}{\bibfnamefont{S.}~\bibnamefont{Jun}}, \bibinfo{journal}{Phys.
  Rev. B} \textbf{\bibinfo{volume}{78}}, \bibinfo{pages}{073405}
  (\bibinfo{year}{2008}).

\bibitem[{\citenamefont{Malola et~al.}(2009)\citenamefont{Malola, H{\"a}kkinen,
  and Koskinen}}]{malola_EPJD_09}
\bibinfo{author}{\bibfnamefont{S.}~\bibnamefont{Malola}},
  \bibinfo{author}{\bibfnamefont{H.}~\bibnamefont{H{\"a}kkinen}},
  \bibnamefont{and} \bibinfo{author}{\bibfnamefont{P.}~\bibnamefont{Koskinen}},
  \bibinfo{journal}{Eur. Phys. J. D} \textbf{\bibinfo{volume}{52}},
  \bibinfo{pages}{71} (\bibinfo{year}{2009}).

\bibitem[{\citenamefont{Huang et~al.}(2009)\citenamefont{Huang, Liu, Su, Wu,
  Duan, Gu, and Liu}}]{bhuang_PRL_09}
\bibinfo{author}{\bibfnamefont{B.}~\bibnamefont{Huang}},
  \bibinfo{author}{\bibfnamefont{M.}~\bibnamefont{Liu}},
  \bibinfo{author}{\bibfnamefont{N.}~\bibnamefont{Su}},
  \bibinfo{author}{\bibfnamefont{J.}~\bibnamefont{Wu}},
  \bibinfo{author}{\bibfnamefont{W.}~\bibnamefont{Duan}},
  \bibinfo{author}{\bibfnamefont{B.}~\bibnamefont{Gu}}, \bibnamefont{and}
  \bibinfo{author}{\bibfnamefont{F.}~\bibnamefont{Liu}},
  \bibinfo{journal}{Phys. Rev. Lett.} \textbf{\bibinfo{volume}{102}},
  \bibinfo{pages}{166404} (\bibinfo{year}{2009}).

\bibitem[{\citenamefont{Jiang et~al.}(2007)\citenamefont{Jiang, Sumpter, and
  Dai}}]{jiang_JCP_07}
\bibinfo{author}{\bibfnamefont{D.}~\bibnamefont{Jiang}},
  \bibinfo{author}{\bibfnamefont{B.~G.} \bibnamefont{Sumpter}},
  \bibnamefont{and} \bibinfo{author}{\bibfnamefont{S.}~\bibnamefont{Dai}},
  \bibinfo{journal}{J. Chem. Phys.} \textbf{\bibinfo{volume}{126}},
  \bibinfo{pages}{134701} (\bibinfo{year}{2007}).

\bibitem[{\citenamefont{Kobayashi et~al.}(2006)\citenamefont{Kobayashi, Fukui,
  Enoki, and Kusakabe}}]{kobayashi_PRB_06}
\bibinfo{author}{\bibfnamefont{Y.}~\bibnamefont{Kobayashi}},
  \bibinfo{author}{\bibfnamefont{K.-I.} \bibnamefont{Fukui}},
  \bibinfo{author}{\bibfnamefont{T.}~\bibnamefont{Enoki}}, \bibnamefont{and}
  \bibinfo{author}{\bibfnamefont{K.}~\bibnamefont{Kusakabe}},
  \bibinfo{journal}{Phys. Rev. B} \textbf{\bibinfo{volume}{73}},
  \bibinfo{pages}{125415} (\bibinfo{year}{2006}).

\bibitem[{\citenamefont{Koskinen et~al.}(2008)\citenamefont{Koskinen, Malola,
  and H\"akkinen}}]{koskinen_PRL_08}
\bibinfo{author}{\bibfnamefont{P.}~\bibnamefont{Koskinen}},
  \bibinfo{author}{\bibfnamefont{S.}~\bibnamefont{Malola}}, \bibnamefont{and}
  \bibinfo{author}{\bibfnamefont{H.}~\bibnamefont{H\"akkinen}},
  \bibinfo{journal}{Phys. Rev. Lett.} \textbf{\bibinfo{volume}{101}},
  \bibinfo{pages}{115502} (\bibinfo{year}{2008}).

\bibitem[{\citenamefont{Wassmann et~al.}(2008)\citenamefont{Wassmann,
  Seitsonen, Saitta, Lazzeri, and Mauri}}]{wassmann_PRL_08}
\bibinfo{author}{\bibfnamefont{T.}~\bibnamefont{Wassmann}},
  \bibinfo{author}{\bibfnamefont{A.~P.} \bibnamefont{Seitsonen}},
  \bibinfo{author}{\bibfnamefont{A.~M.} \bibnamefont{Saitta}},
  \bibinfo{author}{\bibfnamefont{M.}~\bibnamefont{Lazzeri}}, \bibnamefont{and}
  \bibinfo{author}{\bibfnamefont{F.}~\bibnamefont{Mauri}},
  \bibinfo{journal}{Phys. Rev. Lett.} \textbf{\bibinfo{volume}{101}},
  \bibinfo{pages}{096402} (\bibinfo{year}{2008}).

\bibitem[{\citenamefont{Reddy et~al.}(2009)\citenamefont{Reddy,
  Ramasubramaniam, Shenoy, and Zhang}}]{reddy_APL_09}
\bibinfo{author}{\bibfnamefont{C.~D.} \bibnamefont{Reddy}},
  \bibinfo{author}{\bibfnamefont{A.}~\bibnamefont{Ramasubramaniam}},
  \bibinfo{author}{\bibfnamefont{V.~B.} \bibnamefont{Shenoy}},
  \bibnamefont{and} \bibinfo{author}{\bibfnamefont{W.-W.} \bibnamefont{Zhang}},
  \bibinfo{journal}{Appl. Phys. Lett.} \textbf{\bibinfo{volume}{94}},
  \bibinfo{pages}{101904} (\bibinfo{year}{2009}).

\bibitem[{\citenamefont{Kawai et~al.}(2000)\citenamefont{Kawai, Miyamoto,
  Sugino, and Koga}}]{kawai_PRB_00}
\bibinfo{author}{\bibfnamefont{T.}~\bibnamefont{Kawai}},
  \bibinfo{author}{\bibfnamefont{Y.}~\bibnamefont{Miyamoto}},
  \bibinfo{author}{\bibfnamefont{O.}~\bibnamefont{Sugino}}, \bibnamefont{and}
  \bibinfo{author}{\bibfnamefont{Y.}~\bibnamefont{Koga}},
  \bibinfo{journal}{Phys. Rev. B} \textbf{\bibinfo{volume}{62}},
  \bibinfo{pages}{R16349} (\bibinfo{year}{2000}).

\bibitem[{\citenamefont{Girit et~al.}(2009)\citenamefont{Girit, Meyer, Erni,
  Rossell, Kisielowski, Yang, Park, Crommie, Cohen, Louie
  et~al.}}]{girit_science_09}
\bibinfo{author}{\bibfnamefont{{\c{C}}.~{\"O}.} \bibnamefont{Girit}},
  \bibinfo{author}{\bibfnamefont{J.~C.} \bibnamefont{Meyer}},
  \bibinfo{author}{\bibfnamefont{R.}~\bibnamefont{Erni}},
  \bibinfo{author}{\bibfnamefont{M.~D.} \bibnamefont{Rossell}},
  \bibinfo{author}{\bibfnamefont{C.}~\bibnamefont{Kisielowski}},
  \bibinfo{author}{\bibfnamefont{L.}~\bibnamefont{Yang}},
  \bibinfo{author}{\bibfnamefont{C.-H.} \bibnamefont{Park}},
  \bibinfo{author}{\bibfnamefont{M.~F.} \bibnamefont{Crommie}},
  \bibinfo{author}{\bibfnamefont{M.~L.} \bibnamefont{Cohen}},
  \bibinfo{author}{\bibfnamefont{S.~G.} \bibnamefont{Louie}},
  \bibnamefont{et~al.}, \bibinfo{journal}{Science}
  \textbf{\bibinfo{volume}{323}}, \bibinfo{pages}{1705} (\bibinfo{year}{2009}).

\bibitem[{\citenamefont{Shenoy et~al.}(2008)\citenamefont{Shenoy, Reddy,
  Ramasubramaniam, and Zhang}}]{shenoy_PRL_08}
\bibinfo{author}{\bibfnamefont{V.~B.} \bibnamefont{Shenoy}},
  \bibinfo{author}{\bibfnamefont{C.~D.} \bibnamefont{Reddy}},
  \bibinfo{author}{\bibfnamefont{A.}~\bibnamefont{Ramasubramaniam}},
  \bibnamefont{and} \bibinfo{author}{\bibfnamefont{Y.~W.} \bibnamefont{Zhang}},
  \bibinfo{journal}{Phys. Rev. Lett.} \textbf{\bibinfo{volume}{101}},
  \bibinfo{pages}{245501} (\bibinfo{year}{2008}).

\bibitem[{\citenamefont{Lan et~al.}(2009)\citenamefont{Lan, Wand, Gan, and
  Chin}}]{lan_PRB_09}
\bibinfo{author}{\bibfnamefont{J.~H.} \bibnamefont{Lan}},
  \bibinfo{author}{\bibfnamefont{J.-S.} \bibnamefont{Wand}},
  \bibinfo{author}{\bibfnamefont{C.~K.} \bibnamefont{Gan}}, \bibnamefont{and}
  \bibinfo{author}{\bibfnamefont{S.~K.} \bibnamefont{Chin}},
  \bibinfo{journal}{Phys. Rev. B} \textbf{\bibinfo{volume}{79}},
  \bibinfo{pages}{115401} (\bibinfo{year}{2009}).

\end{thebibliography}

\end{document}